\documentclass[aps,pre,twocolumn,groupedaddress,showpacs]{revtex4}
\usepackage{isolatin1}
\usepackage{graphicx}
\usepackage{graphics}
\newcommand{\ud}{\mathrm{d} }
\newcommand{\sgn}{\mathrm{sgn} }
\begin{document}

\title{Effects of additive noise on vibrational resonance in a bistable
system} \author{J. Casado-Pascual} \email{jcasado@us.es}
\affiliation{Universidad de Sevilla, F\'{\i}sica Te\'orica. Apartado de
Correos 1065, Sevilla 41080, Spain} \author{J. P. Baltan\'as} \email{
baltanas@escet.urjc.es} \affiliation{Departamento de Matem\'aticas y 
F\'{\i}sica Aplicadas y Ciencias de la
Naturaleza, Universidad Rey Juan Carlos, Tulip\'an s/n, M\'ostoles
28933, Madrid, Spain} 

\begin{abstract}
We study the overdamped motion of a particle in a bistable potential
subject to the action of a bichromatic force and additive noise, within
the context of the vibrational resonance phenomenon. Under appropriate
conditions, we obtain analytical expressions for the relevant
observables which quantifies this phenomenon. The theoretical results
are compared with those obtained by the numerical solution of the
stochastic differential equation which describes the dynamics of the
system. The limits of validity of the theoretical approach are also
discussed.
\end{abstract}


\maketitle

\section{Introduction}

During the last three decades, a large amount of work has been devoted
to the study of nonlinear systems subject to noise and time-dependent
forces. In the course of these studies it has become clear that, by
contrast to the role played by stochastic forces in linear systems,
noise can drastically alter the response of nonlinear dynamical systems
to the external forcing under some particular circumstances. A
particularly interesting example of the effects of noise within the
framework of signal processing by nonlinear systems is Stochastic
Resonance (SR), i.e., the amplification of a weak input signal by the
concerted actions of noise and the nonlinearity of the system. Although
discussed initially within the context of dynamical systems with
bistable potentials \cite{Benzi1981_jpa}, the phenomenon of SR has also
been found in other dynamical systems
\cite{Gammaitoni1998_rmp,Anishchenko1999_pu}, including systems of
biological interest (see, e.g., Ref.~\cite{Hanggi2002_chphch} and
references therein). Many authors have studied the occurrence of SR not
only in the processing of harmonic signals but also of aperiodic signals
\cite{Collins1995_pre,Collins1996_nature,Chow1998_chaos,Barbay2000_prl},
and in the presence of both white and colored noise
\cite{Neiman1994_prl,Neiman1996_pla,Nozaki1999_prl}. Furthermore,
similar effects have been also found when a chaotic signal is used
instead of noise \cite{Sinha1999_physa}.

The paradigmatic model in the study of SR is the overdamped motion of a
particle in a symmetric double well potential driven by a periodic
signal and noise. Its dynamics is described by the stochastic
differential equation (SDE)
\begin{equation}
\label{lang_sr}
\dot{x}(t)=-U^{\prime}[x(t)]+F(t)+\xi(t),
\end{equation}
where $U'(x)$ is the derivative with respect to $x$ of the symmetric
double well potential $U(x)$, $F(t)$ represents the periodic forcing,
and $\xi(t)$ is a zero average Gaussian white noise with autocorrelation
function $\langle\xi(t)\xi(s)\rangle=2D\delta (t-s)$. SR in this model
can be understood in terms of the synchronization of two time scales:
the Kramers time, characterizing the noise induced interwell
transitions, and the time scale associated to the external driving.
Roughly speaking, when Kramers' time matches half the period of the
external driving for a given noise strength $D$, the amplification of
the weak signal $F(t)$ is optimal.

Recently, an analogous phenomenon named vibrational resonance (VR) has
been shown to occur when the noise is replaced by a high-frequency
periodic force of varying amplitude. Originally described by Landa and
McClintock \cite{Landa2000_jpa}, its study has been also addressed by
other authors and, from different points of view, in excitable
\cite{Ullner2003_pla}, spatially extended \cite{Zaikin2002_pre}, and
bistable systems \cite{Blekhman2004_ijnm,Baltanas2003_pre}. In
Ref.~\cite{Baltanas2003_pre}, a brief numerical study of the effects of
additive noise on VR has been presented, but a more detailed
investigation of this topic is still lacking. In this paper, we
undertake an analytical study of the problem that will provide explicit
expressions for the relevant observables. Our analytical results are
compared with numerical simulations performed on the model described in
Ref.~\cite{Baltanas2003_pre}.

\section{Description of the model and characterization of VR}

\label{desc-mod}

We consider a system described by the SDE (\ref{lang_sr}), with $U(x)$
being the symmetric quartic potential in dimensionless form
\begin{equation}
\label{orig_pot}
U(x)=\frac{x^{4}}{4}-\frac{x^{2}}{2},
\end{equation}
and $F(t)$ the bichromatic force
\begin{equation}
\label{bic_force}
F(t)=A \cos(\Omega \,t)+ N \,\Omega \, r \cos(N \,\Omega \,t+\varphi).
\end{equation}
Here, for reasons that will become clear in the next section, we have
introduced in the second monochromatic force an arbitrary initial
dephasing, $\varphi$. The parameter $N$ is chosen to be a positive
integer, so that $F(t)$ is periodic with the same period $T=2
\pi/\Omega$ as the first monochromatic force. By $r$ we have denoted the
ratio of the amplitude of the second force to its frequency. This ratio
$r$ is assumed to be of the same order as the parameters characterizing
the potential $U(x)$. We are interested in situations in which the
parameters $N \,\Omega \,r $ and $N\,\Omega$ appearing in the second
monochromatic force are much larger than the rest of the parameters in
the problem. In this sense, we will say that $N \,\Omega \, r
\cos(N\,\Omega \,t +\varphi)$ is a strong, high-frequency monochromatic
force. This situation can be formally achieved by taking the limit
$N\rightarrow \infty$, with the ratio $r$ kept fixed.

The corresponding Fokker-Planck equation (FPE) for the probability
density $P(x,t)$ reads
\begin{equation}
\label{FPE1}
\frac{\partial}{\partial t}P(x,t)=\frac{\partial}{\partial x}\Bigg[D
\frac{\partial}{\partial x}+ U'(x)-F(t)\Bigg] P(x,t).
\end{equation}
The analysis of this equation is simplified by making use of two
important theorems: the H-theorem, which ensures the existence of a
unique long time distribution function $P_{\infty}(x,t)$
\cite{Lebowitz,Risken}, and the Floquet theorem, which guarantees that
$P_{\infty}(x,t)$ is periodic in time with the same period, $T$, as the
external force \cite{Jung-Hanggi1}. Henceforth, we will assume that the
relaxation transient stage has ended and, consequently, the long time
distribution, $P_{\infty}(x,t)$, has been reached.

The first moment of the probability distribution can be used to
characterize the system response to the external driving. As a
consequence of the above mentioned theorems, its long time limit,
$\langle x(t) \rangle_{\infty}$, is a periodic function of time with
period $T$ and, therefore, it can be expanded in Fourier series as
follows
\begin{eqnarray}
\label{Fourierseries}
\langle x(t)\rangle_{\infty}&=&\sum_{n=-\infty}^{\infty} X_{n}\,
e^{i \,n \,\Omega \,t}\nonumber \\ &=& 
X_0+2\sum_{n=1}^{\infty}|X_{n}| \cos\left(n \,\Omega\,
t-\phi_{n}\right),
\end{eqnarray}
where 
\begin{equation}
\label{Fouriercoeff}
X_{n}=\frac{1}{T} \int_{0}^{T} \ud t\,\langle x(t)\rangle_{\infty}\,
e^{-i\, n \,\Omega \,t},
\end{equation}
and the phases $\phi_{n}$ have been chosen so that
\begin{eqnarray}
\label{phases}
\cos \phi_{n}&=&\mathrm{Re}\left(\frac{ X_{n}}{| X_{n}|}\right), \\ \sin
\phi_{n} &=&-\mathrm{Im}\left(\frac{ X_{n}}{|X_{n}|}\right).
\end{eqnarray} 

The observable that has been usually considered to quantify the VR
phenomenon is the parameter
\begin{equation}
\label{response}
Q=2\frac{|X_1|}{A}.
\end{equation}
This parameter represents the ratio of the amplitude associated to the
first harmonic to $A$, and it is directly related to the spectral
amplification $\eta$ through the expression $Q=\sqrt{\eta}$ (see
Ref.~\cite{Gammaitoni1998_rmp}). Notice that the above definition of the
parameter $Q$ differs by a factor $2$ from the definition used in the
theoretical approach in Ref.~\cite{Baltanas2003_pre}, and by a factor
$A^{-1}$ from the one used in the numerical results appearing in the
same reference. In the deterministic case, $D=0$, it has been shown that
$Q$ goes through a maximum as the intensity of the high-frequency force
is increased. This maximum is approximately localized around the value
$r=\sqrt{2/3}$ of the ratio of the amplitude of the second force to its
frequency \cite{Blekhman2004_ijnm,Baltanas2003_pre}. This behavior
resembles that of SR, with the noise strength $D$ playing the role of
the intensity of the high-frequency force.

When white noise is added, the situation is different.  More precisely,
the two main effects of increasing the value of the noise strength $D$
are that the value at the maximum of the resonance curve decreases and,
at the same time, its location is shifted towards lower values of the
high-frequency amplitude. Even more, for large enough values of $D$, the
VR phenomenon completely disappears \cite{Baltanas2003_pre}. These
effects can be qualitatively understood as a consequence of the fact
that the white noise provides an input to the system with contributions
to all the frequencies. On the one hand, the fraction corresponding to
the high-frequency region advances the appearance of the maximum. On the
other hand, the remaining contribution masks the high-frequency force,
decreasing its relative importance \cite{Baltanas2003_pre}. To shed some
more light on this question, an extension of the theoretical approach
put forward in Ref.~\cite{Baltanas2003_pre} to the case in which white
noise is present would be desirable. This is the main aim of the next
section.

\section{Theoretical approach}

In this section, we will obtain analytical expressions for the parameter
$Q$ based on three approximations of different nature. The first two
ones are simply generalizations to the noisy problem of those carried
out in the theoretical approach developed in the absence of noise
\cite{Baltanas2003_pre}. The third approximation is specific to the
noisy problem. For clarity in the exposition, as well as in the
discussion of their validity conditions, we will present them
separately.

\subsection{Derivation of the effective dynamics}
\label{EffDyn}
As a consequence of Eqs.~(\ref{bic_force}) and (\ref{FPE1}), the time
derivative of the probability density $P(x,t)$ diverges as $N$ in the
limit $N\rightarrow \infty$. To avoid this divergence, it is convenient
to extract the {\it fast} dependence from $x(t)$, and define the new
stochastic process
\begin{equation}
\label{slow_var}
y(t)=x(t)-r \sin\left(N\,\Omega \,t+\varphi\right).
\end{equation}
The Fourier components of $\langle x(t) \rangle_{\infty}$
and $\langle y(t) \rangle_{\infty}$ are related according to
\begin{equation}
\label{relFoucoeff}
Y_n=X_n+\frac{i r}{2}\left(e^{i \varphi} \delta_{n,N}- e^{-i \varphi}
\delta_{n,-N}\right),
\end{equation}
where $Y_n$ is defined from Eq.~(\ref{Fouriercoeff}), by replacing
$\langle x(t) \rangle_{\infty}$ by $\langle y(t)
\rangle_{\infty}$. Therefore, as $N\neq\pm 1$, we can substitute the
coefficient $X_1$ by $Y_1$ in the definition of the parameter $Q$ [see
Eq.~(\ref{response})].

The advantage of using the process $y(t)$ instead of $x(t)$ becomes
clear after writing the FPE for its probability density, which reads
\begin{eqnarray}
\label{FPE2}
\frac{\partial}{\partial t}{\mathcal P}(y,t,\varphi)&=&
\frac{\partial}{\partial y}\Bigg[D \frac{\partial}{\partial y}+
{\mathcal U}'\left(y,t,\varphi\right) \nonumber \\ &-&A \cos(\Omega\,
t)\Bigg]{\mathcal P}(y,t,\varphi).
\end{eqnarray}
Here, we have introduced the new potential ${\mathcal
U}(y,t,\varphi):=U[y+r \sin(N\,\Omega\,t+\varphi)]$, and the dependence
of the functions on the phase $\varphi$ has been written
explicitely.  From Eq.~(\ref{FPE2}) and the definition of ${\mathcal
U}(y,t,\varphi)$ it follows that the time derivative of ${\mathcal
P}(y,t,\varphi)$ is at most of order $1$ as $N\rightarrow
\infty$. Therefore, a large number of oscillations of the function $r
\sin(N\,\Omega\,t+\varphi)$ appearing in the definition of ${\mathcal
U}(y,t,\varphi)$ take place before a significant change in ${\mathcal
P}(y,t,\varphi)$ occurs. As a consequence, it is to be expected that,
for $N\gg 1$, ${\mathcal P}(y,t,\varphi)$ is almost independent of the
phase $\varphi$. If we define the phase average of an arbitrary
function $f(y,t,\varphi)$ as
\begin{equation}
\label{defvar}
\overline{f} (y,t)=\overline{f (y,t,\varphi)}:=\frac{1}{2\pi}
\int_{0}^{2 \pi} \ud \varphi \;f(y,t,\varphi),
\end{equation}
then, for $N \gg 1$ and any value of $\varphi\in [0,2\pi ]$, one can
approximate $\mathcal{P}(y,t,\varphi)\approx \overline{{\cal
P}}(y,t)$. Furthermore, the decoupling approximation
$\overline{f(y,t,\varphi) {\cal P}(y,t,\varphi)}\approx
\overline{f}(y,t) \;\overline{{\cal P}}(y,t)$ also holds. Carrying out
the phase average in Eq.~(\ref{FPE2}) and using the decoupling
approximation, one obtains 
\begin{equation}
\label{FPE3}
\frac{\partial}{\partial t}{\overline {\cal
P}}(y,t)=\frac{\partial}{\partial y}\Bigg[D \frac{\partial}{\partial y}+
U'_{eff}(y)\nonumber -A \cos(\Omega \,t)\Bigg]{\overline {\cal P}}(y,t),
\end{equation}
where we have introduced the effective potential
$U_{eff}(y):=\overline{{\mathcal U}(y,t,\varphi)}$, which is clearly
time independent. This averaging procedure has been previously used in
the study of thermal activation in bistable systems under the influence
of a periodic force with moderate to large frequency \cite{Jung89}. It
has also been used in the explanation of the enhancement observed in the
tunnel splitting of a quantum bistable system when a very high frequency
driving is applied \cite{Grossmann}.

It also follows from the above considerations that, for $N\gg 1$,
$\langle y(t) \rangle_{\infty}\approx\langle y(t)
\rangle_{\infty}^{(eff)} :=\int_{-\infty}^{\infty} \ud y \, y
\,{\overline {\cal P}}_{\infty}(y,t)$, where ${\overline {\cal
P}}_{\infty}(y,t)$ is the long time distribution corresponding to the
FPE (\ref{FPE3}). In particular, within this approximation, the
parameter $Q$ is given by
\begin{equation}
\label{Qapp1}
Q=2\frac{\big|Y_{1}^{(eff)}\big|}{A},
\end{equation} 
where $Y_{1}^{(eff)}$ is obtained from Eq.~(\ref{Fouriercoeff}) with
$n=1$, by replacing $\langle x(t) \rangle_{\infty}$ by $\langle y(t)
\rangle_{\infty}^{(eff)}$.

The explicit calculation of $U_{eff}(y)$ for the potential in
Eq.~(\ref{orig_pot}) leads to
\begin{equation}
\label{U_eff}
U_{eff}(y)=\frac{y^{4}}{4}- a(r)\frac{y^{2}}{2},
\end{equation}
where we have dropped an irrelevant constant and have introduced the
quantity $a(r)=1-3 r^2/2$. From this result, it is clear that the
stability of this effective potential depends on the ratio $r$. More
precisely, if $r<\sqrt{2/3}$ the potential is bistable, whereas if
$r\geq\sqrt{2/3}$ it is monostable. Thus, an increase in $r$ leads to a
decrease in the effective barrier height and, eventually, to its
disappearance.

In summary, we have shown that, for $N\gg 1$, the time evolution of the
original stochastic process $x(t)$ can be approximately described in
terms of the dynamics of a Brownian particle moving in the effective
potential $U_{eff}(y)$ and in the presence of noise and the
monochromatic force $A\cos(\Omega\,t)$.

\subsection{Linear Response Theory}

Henceforth, following the approach in Ref.~\cite{Baltanas2003_pre}, we
will assume that the amplitude $A$ is small enough, so that Linear
Response Theory (LRT) provides a good description of the effective
dynamics obtained in the previous subsection.  An extensive study of the
validity conditions of LRT can be found in
Refs.~\cite{jcasado1,jcasado2,jcasado3}. Within LRT the parameter $Q$ in
Eq.~(\ref{Qapp1}) is given by \cite{Gammaitoni1998_rmp}
\begin{equation}
\label{QLRT}
Q=\left|{\widehat\chi}_{eff}(\Omega)\right|,
\end{equation}
where 
\begin{equation}
\label{FT}
{\widehat\chi}_{eff}(\Omega)=\int_{0}^{\infty} \ud t\,\chi_{eff}(t)\,
 e^{-i\Omega\, t}
\end{equation}
is the value at $\omega=\Omega$ of the Fourier transform of the
response function, $\chi_{eff}(t)$, corresponding to an overdamped
Brownian particle moving in the effective potential $U_{eff}(y)$. This
response function obeys the well-known Fluctuation-Dissipation Theorem
(FDT) \cite{Kubo1,Kubo2,Thomas}
\begin{equation}
\label{FDT}
\chi_{eff}(t)=-\frac{\eta(t)}{D}\frac{\ud}{\ud t}K_{eff}(t),
\end{equation}
where $\eta(t)$ is the Heaviside step function, and $K_{eff}(t)=\langle
y(t) y(0)\rangle_{eq}^{(eff)}$ is the equilibrium autocorrelation
function of the effective system described by the FPE (\ref{FPE3}) in
the absence of external driving. Thus, the stochastic process $y(t)$
appearing in the definition of $K_{eff}(t)$ is a solution of the SDE
\begin{equation}
\label{SDEy}
\dot{y}(t)=-U_{eff}^{\prime}[y(t)]+\xi(t).
\end{equation}
Use of Eq.~(\ref{FDT}) in Eq.~(\ref{FT}) yields
\begin{eqnarray}
\label{Rechi}
\mathrm{Re}\left[{\widehat\chi}_{eff}(\Omega)\right]
&=&-\frac{\Omega}{D}\Bigg[\int_0^{\infty} \ud t \, K_{eff}(t) \sin
\left(\Omega\, t \right)\nonumber \\
&&-\frac{K_{eff}(0)}{\Omega}\Bigg],\\
\label{Imchi}
\mathrm{Im}\left[{\widehat \chi}_{eff}(\Omega)\right]
&=&-\frac{\Omega}{D} \int_0^{\infty} \ud t \, K_{eff}(t) \cos
\left(\Omega\, t \right).
\end{eqnarray}
These two equations, together with Eq.~(\ref{QLRT}), allow us to express
$Q^{(LRT)}$ in terms of the equilibrium autocorrelation function
$K_{eff}(t)$.

For the nonlinear potential in Eq.~(\ref{U_eff}), explicit expressions
for $K_{eff}(t)$ are unknown, so that this correlation function must be
evaluated either numerically or by resorting to suitable
approximations. Before applying to our problem approximate techniques
discussed in the literature, it is convenient to reduce the SDE
(\ref{SDEy}) to a more standard form. In order to do so, let us rescale
the coordinate and time, and define the new stochastic process
$\tilde{y}\big(\tilde{t}\big)=|a(r)|^{-1/2} y\big(|a(r)|^{-1} {\tilde
t}\big)$ (for $r\neq\sqrt{2/3}$).  Then, from Eq.~(\ref{SDEy}), it is
easy to verify that $\tilde{y}\left(\tilde{t}\right)$ fulfills the SDE
\begin{equation} 
\label{SDEytilde0}
\dot{{\tilde y}}({\tilde t})=-{\widetilde U}^{\prime}[{\tilde y}({\tilde
t})]+{\tilde \xi}({\tilde t}),
\end{equation}
where we have introduced the rescaled potential 
\begin{equation}
\label{rescaledU}
{\widetilde U}\big({\tilde y}\big)=\frac{{\tilde y}^4}{4}
+\sgn\big[a(r)\big]\frac{{\tilde y}^2}{2},
\end{equation}
and the rescaled Gaussian white noise ${\tilde \xi}\big({\tilde t}\big)$
with zero average and autocorrelation function $\langle {\tilde
\xi}\big({\tilde t}\big){\tilde \xi}\big({\tilde
t^{\prime}}\big)\rangle=2\, |a(r)|^{-2}\, D \,\delta\big({\tilde
t}-{\tilde t}^{\prime}\big)$. From these considerations, it is
straightforward to prove that the equilibrium autocorrelation function
of the original process, $K_{eff}(t,D)$, and that of the rescaled one,
${\widetilde K}({\tilde t},{\widetilde D})$, are related by
\begin{equation}
\label{krelation}
K_{eff}(t,D)=|a(r)| {\widetilde K}\left(|a(r)| t,\frac{D}{|a(r)|^{2}}
\right).
\end{equation}
Analogously, Eqs.~(\ref{Rechi}), (\ref{Imchi}) and (\ref{krelation})
lead to the following relation between the Fourier transform of the
response function of the original process, $\big|{\widehat
\chi}_{eff}(\Omega,D)\big|$, and that of the rescaled one,
$\big|{\widehat{\widetilde \chi}}({\widetilde \Omega},{\widetilde
D})\big|$,
\begin{equation}
\label{chirelation}
\left|{\widehat \chi}_{eff}(\Omega,D)\right|=\frac{1}{|a(r)|}
\left|\widehat{{\widetilde
\chi}}\left(\frac{\Omega}{|a(r)|},\frac{D}{|a(r)|^{2}} \right)\right|.
\end{equation}

\subsection{Weak Noise Approximation}
\label{WNA}
The expressions (\ref{krelation}) and (\ref{chirelation}) allow us to
evaluate $K_{eff}(t,D)$ and $\big|{\widehat \chi}_{eff}(\Omega,D)\big|$
from the rescaled functions ${\widetilde K}\big({\tilde t},{\widetilde
D}\big)$ and $\big|\widehat{{\widetilde \chi}}\big({\widetilde
\Omega},{\widetilde D}\big)\big|$. The asymptotic behaviors of these
rescaled functions for small values of the noise strength ${\widetilde
D}$ have been widely studied in the literature. To summarize the
results, we will consider separately two cases.

Firstly, if $r<\sqrt{2/3}$, then ${\widetilde U}({\tilde y})={\tilde
y}^4/4-{\tilde y}^2/2$ is the archetypal symmetric quartic double-well
potential expressed in dimensionless form. For this bistable potential,
we can use the two-mode approximation \cite{Jung-Hanggi2}. This
approximation is based on the existence of a clear-cut separation
between the time scales associated to inter-well and intra-well motions,
and it is expected to be valid for small values of the noise strength
${\widetilde D}$. Within this approximation, ${\widetilde K}\big({\tilde
t},{\widetilde D}\big)$ and $\big|\widehat{{\widetilde
\chi}}\big({\widetilde \Omega},{\widetilde D}\big)\big|$ are given by
\begin{equation}
\label{ACF1}
{\widetilde K}\big({\tilde t},{\widetilde D}\big)=g_1\big({\widetilde
D}\big) \exp\big[- \lambda_1\big({\widetilde D}\big)\,{\tilde
t}\big]+g_2\big({\widetilde D}\big) \exp\big[-\alpha \,{\tilde t}\big],
\end{equation}
and
\begin{eqnarray}
\label{chi1}
\big|\widehat{{\widetilde \chi}}\big({\widetilde \Omega},{\widetilde
D}\big)\big|\!\!&=&\!\!\frac{1}{{\widetilde D}}\Bigg\{\frac{2 \alpha
g_1\big({\widetilde D}\big)g_2\big({\widetilde D}\big)
\lambda_1\big({\widetilde D}\big)\big[\alpha \lambda_1\big({\widetilde
D}\big)+{\widetilde \Omega}^2 \big] }{\big[\lambda_1^{2}\big({\widetilde
D}\big)+{\widetilde \Omega}^2\big]\big(\alpha^2+{\widetilde
\Omega}^2\big)} \nonumber\\ &&\!\!+\frac{g_1^{2}\big({\widetilde
D}\big)\lambda_1^{2}\big({\widetilde
D}\big)}{\lambda_1^{2}\big({\widetilde D}\big)+{\widetilde \Omega}^2}
+\frac{g_2^{2}\big({\widetilde D}\big)\alpha^2}{\alpha^2+{\widetilde
\Omega}^2}\Bigg\}^{1/2}.
\end{eqnarray}
In the above expressions, the parameter $\alpha={\widetilde U}^{\prime
\prime}\big(\pm 1\big)=2$ is the curvature at the minima of the
potential ${\widetilde U}\big({\tilde y}\big)$, and
$\lambda_1({\widetilde D})$ is the the smallest non-vanishing eigenvalue
of the Fokker-Planck operator corresponding to the SDE
(\ref{SDEytilde0}). In the steepest-descent approximation, this
eigenvalue is given to leading order in ${\widetilde D}$ by
\begin{equation}
\label{lamb1}
\lambda_1\big({\widetilde D}\big)\approx\frac{\sqrt{2}}{\pi}
\exp\left(-\frac{1}{4{\widetilde D}} \right).
\end{equation}
The weights $g_1\big({\widetilde D}\big)$ and $g_2\big({\widetilde
D}\big)$ read
\begin{equation}
\label{g2}
g_2\big({\widetilde D}\big)=\frac{\big[\lambda_1\big({\widetilde
D}\big)+1\big] m_2\big({\widetilde D}\big)-m_4\big({\widetilde
D}\big)}{\lambda_1\big({\widetilde D}\big)-\alpha},
\end{equation}
\begin{equation}
\label{g1}
g_1\big({\widetilde D}\big)=m_2\big({\widetilde
D}\big)-g_2\big({\widetilde D}\big),
\end{equation}
where
\begin{eqnarray}
\label{moments}
m_{2 n}\big({\widetilde D}\big)&=&\frac{\int_{-\infty}^{\infty} \ud
{\tilde y}\, {\tilde y}^{2 n}\,e^{-{\widetilde U}\big({\tilde
y}\big)/\widetilde D}}{\int_{-\infty}^{\infty} \ud {\tilde y}\,
e^{-{\widetilde U}\big({\tilde y}\big)/{\widetilde D}}}\nonumber \\
&=&\big(2{\widetilde D}\big)^{n/2} \,\frac{\Gamma(n+1/2)
D_{-n-1/2}\big[-\big(2{\widetilde D} \big)^{1/2}\big]}{\Gamma(1/2)
D_{-1/2}\big[-\big(2{\widetilde D} \big)^{1/2}\big]}\nonumber \\
\end{eqnarray}
are the nonvanishing moments of the equilibrium distribution. Using the
asymptotic expansion of the parabolic cylinder functions for small
values of ${\widetilde D}$, these weights can be approximated by
$g_1\big({\widetilde D}\big)\approx 1-\left(\alpha+1\right){\widetilde
D}/\alpha$ and $g_2\big({\widetilde D}\big)\approx {\widetilde
D}/\alpha$ to first order in ${\widetilde D}$.

Secondly, if $r>\sqrt{2/3}$, then ${\widetilde U}({\tilde y})={\tilde
y}^4/4+{\tilde y}^2/2$ is a monostable potential. In this case, for
small values of ${\widetilde D}$, the autocorrelation function
${\widetilde K}\big({\tilde t},{\widetilde D}\big)$ is given by
\cite{Gardiner}
\begin{equation}
\label{ACF2}
{\widetilde K}\big({\tilde t},{\widetilde D}\big)={\widetilde D} \,\exp
\big[-\big|\Lambda_1\big({\widetilde D}\big)\big|\,{\tilde t} \big],
\end{equation}
with
\begin{equation}
\label{lamb1new}
\Lambda_1\big({\widetilde D}\big)=-2+\big(1-6 {\widetilde D}\big)^{1/2}.
\end{equation}
Consequently, the Fourier transform of the response function reads
\begin{equation}
\label{chi2}
\big|\widehat{{\widetilde \chi}}\big({\widetilde \Omega},{\widetilde
D}\big)\big|=\frac{\big| \Lambda_1\big({\widetilde
D}\big)\big|}{\big[\Lambda_1^2\big({\widetilde D}\big)+{\widetilde
\Omega}^2\big]^{1/2}}.
\end{equation}
To evaluate $K_{eff}(t,D)$ and $\big|{\widehat
\chi}_{eff}(\Omega,D)\big|$ one simply introduces either the expressions
(\ref{ACF1}) and (\ref{chi1}) or (\ref{ACF2}) and (\ref{chi2}) into
Eqs.~(\ref{krelation}) and (\ref{chirelation}). Finally, within this
approximation, the parameter $Q$ will be evaluated by replacing in
Eq.~(\ref{QLRT}) the result obtained for $\big|{\widehat
\chi}_{eff}(\Omega,D)\big|$.  It is important to emphasize that the
approximate expressions (\ref{ACF1}), (\ref{chi1}), (\ref{ACF2}) and
(\ref{chi2}) have been obtained under the assumption that ${\widetilde
D}$ is a small parameter. Therefore, the resulting expressions for
$K_{eff}(t,D)$, $\big|{\widehat \chi}_{eff}(\Omega,D)\big|$ and $Q$ are
expected to be valid only for small values of $D\,|a(r)|^{-2}$. Taking
into account that, for $D\neq 0$, the parameter $D\,|a(r)|^{-2}$
diverges at $r= \sqrt{2/3}$, then it is clear that, even for $D\ll
1$, there exists a region around $r= \sqrt{2/3}$ in which the resulting
expressions for $K_{eff}(t,D)$, $\big|{\widehat
\chi}_{eff}(\Omega,D)\big|$ and $Q$ are not applicable. In particular,
the expression for $Q$ is expected to provide a successful explanation
of the VR phenomenon whenever its maximum takes place at a value of $r$
outside the above mentioned region.

From the above considerations, it is also possible to obtain an
expression for the location of the maximum characterizing the VR
phenomenon in the presence of white noise. According to the numerical
results put forward in Ref.~\cite{Baltanas2003_pre}, this maximum is
located at a value $r_M$ such that $r_M<\sqrt{2/3}$. Then, to evaluate
the parameter $Q$ in the neighborhood of $r_M$, one has to consider
Eq.~(\ref{chi1}) and follow the procedure described in the above
paragraph. As the expression thus obtained is rather cumbersome, it is
convenient to introduce a further approximation. More precisely, we will
assume that, around $r_M$, the contribution due to the intrawell motion
is negligible, so that just the second term in Eq.~(\ref{chi1}) with
$g_1({\widetilde D})=1$ is present. Therefore, the parameter $Q$ as a
function of $r$ obtained from Eqs.~(\ref{QLRT}) and (\ref{chirelation})
is given by
\begin{equation}
\label{withointrwell}
Q(r)=\frac{a(r)}{D}
\frac{\lambda_1\left[D/a^{2}(r)\right]}{\left\{\lambda_1^2
\left[D/a^{2}(r)\right]+\left[\Omega/a(r)\right]^2
\right\}^{1/2}},
\end{equation}
for $r<\sqrt{2/3}$. The location of the maximum, $r_M$, is
readily determined from the equation $Q^{\prime}(r_M)=0$, where
$Q^{\prime}(r)$ represents the derivative of $Q(r)$ with respect to
$r$. Consequently, from Eq.~(\ref{withointrwell}) it follows that $r_M$
obeys the transcendental equation
\begin{equation}
\label{transceq}
\lambda_1^2\left[\frac{D}{a^2(r_M)}\right]=2 \left[\frac{\Omega}{a(r_M)}
\right]^2 \left[\frac{a^2(r_M)}{4 D}-1\right].
\end{equation}
After making the replacements $D/a^2(r_M) \rightarrow D$ and
$\Omega/a(r_M) \rightarrow \Omega$ in the above equation, the resulting
expression resembles, except for a factor of $2$, the transcendental
equation obtained within the two-state model to determine the noise
intensity that maximizes the spectral amplification (see, e.g.,
Ref.~\cite{Gammaitoni1998_rmp} and references therein). The difference
in the factor of $2$ is due to the fact that, in that case, only the
noise intensity is varied, whereas in our case, by varying $r$, we are
simultaneously modifying the frequency and the noise intensity.

The transcendental equation (\ref{transceq}) provides an explanation for
the numerically observed behavior of the location of the maximum as a
function of $D$ (see Sec.~\ref{desc-mod}). For a fixed value of $D$, the
function $\lambda_1^2\left[D/a^2(r)\right]$ is a monotonously increasing
function of $r$ in the interval $[0,\sqrt{2/3})$. Also, for fixed values
of $D$ and $\Omega$, $2 \left[\Omega/a(r) \right]^2 \left[a^2(r)/(4
D)-1\right]$ is a monotonously decreasing function of $r$ in the same
interval. Thus, the existence of a solution of Eq.~(\ref{transceq}) in
the interval $(0,\sqrt{2/3})$ requires the inequality
$\lambda_1^2\left[D/a^2(0)\right]=\lambda_1^2\left(D\right)<2
\left[\Omega/a(0) \right]^2 \left[a^2(0)/(4
D)-1\right]=2\,\Omega^2\left[1/(4 D)-1\right]$. Therefore, the same
inequality is required for the existence of VR. If we now keep fixed $r$
and $\Omega$ and look at both sides of Eq.~(\ref{transceq}) as a
function of $D$, we notice that $2 \left[\Omega/a(r) \right]^2
\left[a^2(r)/(4 D)-1\right]$ is a monotonously decreasing function of
$D$, whereas $\lambda_1^2\left[D/a^2(r)\right]$ is a monotonously
increasing function of $D$. We then conclude that the abscissa at the
intersection point between these two functions, $r_M$, is shifted
towards the left as $D$ increases. This is exactly the behavior observed
numerically in Ref.~\cite{Baltanas2003_pre}. Obviously, VR disappears
for values of $D$ greater than the critical value $D_c$ for which
$r_M=0$. This critical value fulfills the transcendental equation
\begin{equation}
\label{D_c}
\lambda_1^2\left(D_c\right)=2 \,\Omega^2 \left[\frac{1}{4 D_c}-1\right].
\end{equation}

\section{Comparison with the numerical solution}
In this section, we compare the results obtained for the parameter $Q$
by means of the numerical solution of the SDE (\ref{lang_sr}), with the
analytical expressions obtained in the previous section. In order to
evaluate numerically the parameter $Q$, a large enough number of
stochastic trajectories, $x^{(k)}(t)$, have been generated by
integrating the SDE (\ref{lang_sr}) for every realization of the white
noise $\xi(t)$, starting from a given initial condition $x_0$.  The
numerical solution of the SDE (\ref{lang_sr}) has been carried out by
using the algorithm put forward in Ref.~\cite{Greenside} (see also the
Appendix in Ref.~\cite{jcasado3}). After allowing for a relaxation
transient stage, we evaluate the long time average by
\begin{equation}
\label{numerical}
\langle x(t)\rangle_{\infty}=\frac{1}{M} \sum_{k=1}^{M} x^{(k)}(t),
\end{equation}
where $M$ is the number of stochastic trajectories considered. Then, the
parameter $Q$ is obtained from Eqs.~(\ref{response}) and
(\ref{Fouriercoeff}), with $n=1$, by numerical quadrature.

\begin{figure}
\begin{center}
\scalebox{0.45}{\includegraphics{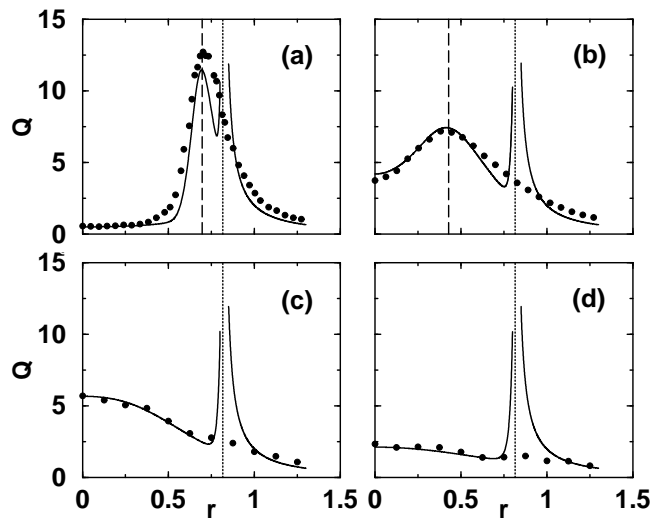}}
\end{center}
\caption{
\label{comparison} 
Dependence of the parameter $Q$ {\it vs}. $r$ corresponding to the noise
strength values $D=1.34\times 10^{-2}$ [panel (a)], $D=6.72 \times
10^{-2}$ [panel (b)], $D=1.34 \times 10^{-1}$ [panel (c)], and
$D=4.03\times10^{-1}$ [panel (d)]. The rest of the parameter values are:
$N=50$, $\varphi=0$, $\Omega=3.4\times 10^{-2}$, $A=1.1\times
10^{-2}$. With circles we have plotted the results obtained from the
numerical solution of the SDE (\ref{lang_sr}) together with
Eqs.~(\ref{orig_pot}) and (\ref{bic_force}).  With solid lines we have
depicted the analytical values of $Q$ provided by Eqs.~(\ref{QLRT}),
(\ref{chirelation}) and either Eqs.~(\ref{chi1})-(\ref{moments}), for
$r<\sqrt{2/3}$, or Eqs.~(\ref{chi2}) and (\ref{lamb1new}), for
$r>\sqrt{2/3}$. The dotted lines indicate the location of the critical
value $r=\sqrt{2/3}$, and the dashed lines the location of the maximum
predicted by Eq.~(\ref{transceq}).}
\end{figure}

In Fig.~\ref{comparison}, we depict the dependence of $Q$ {\it vs}. $r$.
The following set of parameter values has been considered: $N=50$,
$\varphi=0$, $\Omega=3.4\times 10^{-2}$, $A=1.1\times 10^{-2}$,
$D=1.34\times 10^{-2}$ in panel (a), $D=6.72 \times10^{-2}$ in panel
(b), $D=1.34 \times10^{-1}$ in panel (c), and $D=4.03\times10^{-1}$ in
panel (d). They correspond to four representative cases appearing in
Fig.~6 in Ref.~\cite{Baltanas2003_pre}. Notice that the parameter values
appearing in that reference must be rescaled in order to reduce the
bistable potential used there to the standard dimesionless form in
Eq.~(\ref{orig_pot}). With circles we have plotted the results obtained
from the numerical solution of the SDE (\ref{lang_sr}) together with
Eqs.~(\ref{orig_pot}) and (\ref{bic_force}). With solid lines we have
depicted the analytical values of $Q$ provided by Eqs.~(\ref{QLRT}),
(\ref{chirelation}) and either Eqs.~(\ref{chi1})-(\ref{moments}), for
$r<\sqrt{2/3}$, or Eqs.~(\ref{chi2}) and (\ref{lamb1new}), for
$r>\sqrt{2/3}$. The location of the critical value $r=\sqrt{2/3}$ is
indicated by the vertical dotted lines. In addition, in panels (a) and
(b), we have depicted with vertical dashed lines the locations of the
maxima predicted by Eq.~(\ref{transceq}). The critical value of the
noise strength at which the VR phenomenon disappears, obtained from
Eq.~(\ref{D_c}), is $D_c=1.154\times 10^{-1}$. The noise strength values
in panels (c) and (d) are greater than $D_c$, so it is to be expected
that the VR phenomenon is not present in these cases, as it is confirmed
numerically.

A glance to Fig.~1 reveals that the analytical expressions provide a
good description of the main features of the VR phenomenon in the
presence of white noise, at least if the ratio $r$ is far enough away
from the critical value $r=\sqrt{2/3}$. In particular, the analytical
results describe correctly the shift of the location of the maximum
towards lower values of $r$, as well as the decrease of the value of $Q$
at the maximum, as the noise strength increases. It is important to
emphasize that, as we mentioned in Sec.~\ref{WNA}, the analytical
results are not applicable in a neighborhood of the critical value
$r=\sqrt{2/3}$, as a consequence of the divergence of the effective
noise strength $D/[a(r)]^2$. The appearance of the vertical asymptote at
this critical value makes this failure evident. Notice that, in panel
(a), the agreement between the numerical and analytical results around
the maximum gets worse. This is due to the proximity of the location of
this maximum to the critical value $r=\sqrt{2/3}$.

\section{Conclusions}

We have studied the motion of a particle in a bistable potential in the
presence of white noise and an external bichromatic force, within the
context of the VR phenomenon. Analytical expressions for the parameter
$Q$ which quantifies this phenomenon have been obtained based on three
simplifying approximations of different nature. Firstly, the exact
dynamics of the system has been reduced to that of a Brownian particle
moving in an effective potential and under the influence of white noise
and one of the monochromatic forces. In order to make this
approximation, we have assumed that the other monochromatic force is a
strong, high-frequency field. Secondly, we have applied LRT to
express the parameter $Q$ in terms of the equilibrium autocorrelation
function associated to the effective dynamics in the absence of the
external driving. Finally, we have obtained analytical expressions for
$Q$ within the weak noise approximation. To do so, it has been necessary
to consider two disjoint regions of values of the ratio $r$ separated by
the critical value $r=\sqrt{2/3}$. We have also obtained a
transcendental equation for the location of the maximum of $Q(r)$ which
characterizes the VR phenomenon, as well as for the condition under
which this phenomenon dissapears. Comparison with the numerical solution
of the original SDE shows that these analytical expressions provide a
good quantitative description of the VR phenomenon, at least if it takes
place outside a critical region around $r=\sqrt{2/3}$. With this work,
we have attempted to provide a theoretical complement to the numerical
study of the influence of additive noise on VR carried out in
Ref.~\cite{Baltanas2003_pre}.

\begin{acknowledgments}
We acknowledge the support of the Dirección General de Enseñanza
Superior of Spain by the projects BFM2002-03822 (J. C.-P.) and
BFM2000-0967 (J. P. B.). We also want to thank Prof. M. Morillo for his
critical reading of the manuscript and fruitful discussions.
\end{acknowledgments}

\end{document}